\documentclass[twocolumn,secnumarabic,amssymb, nobibnotes, aps, prd,nofootinbib,superscriptaddress]{revtex4-2}

\usepackage[T1]{fontenc} 

\usepackage[T1]{fontenc} 
\usepackage[utf8]{inputenc} 
\usepackage{microtype} 
\usepackage{lmodern} 
\usepackage{booktabs} 
\usepackage{mdframed} 
\usepackage{graphicx}

\usepackage[backgroundcolor=white]{todonotes}

\usepackage{hyperref}
 \usepackage{color}
 \definecolor{dark-red}{rgb}{0.4,0.15,0.15}
 \definecolor{dark-blue}{rgb}{0.15,0.15,0.4}
 \definecolor{medium-blue}{rgb}{0,0,0.5}
 \hypersetup{colorlinks, linkcolor={dark-red}, citecolor={dark-blue}, urlcolor={medium-blue}}

\usepackage{amsmath,amsfonts,amssymb}
\usepackage{eucal} 
\usepackage{mleftright} \mleftright
\usepackage{tensor} 
\usepackage{braket} 
\usepackage{bm} 

\usepackage{mathrsfs} 

\providecommand*{\dd}{\mathop{}\!d}
\renewcommand*{\dd}{\mathop{}\!d}

\providecommand*{\pd}{\mathop{}\!\partial}
\renewcommand*{\pd}{\mathop{}\!\partial}
\providecommand*{\vd}{\mathop{}\!\delta}
\renewcommand*{\vd}{\mathop{}\!\delta}
\providecommand*{\cd}{\mathop{}\!\nabla}
\renewcommand*{\cd}{\mathop{}\!\nabla}

\providecommand*{\Ld}{\mathsterling} 
\renewcommand*{\Ld}{\mathsterling}

\providecommand*{\R}{{\mathbb{R}}}
\renewcommand*{\R}{{\mathbb{R}}}

\begin{document}

\title{Asymptotic symmetries and soft charges of fractons}%

\author{Alfredo Pérez}%
\email{alfredo.perez@uss.cl}
\affiliation{Centro de Estudios Científicos (CECs), Avenida Arturo Prat 514, Valdivia, Chile}
\affiliation{Facultad de Ingeniería, Arquitectura y Diseño, Universidad San Sebastián, sede Valdivia, General Lagos 1163, Valdivia 5110693, Chile}
\author{Stefan Prohazka}%
\email{stefan.prohazka@ed.uk.ac}
\affiliation{School of Mathematics and Maxwell Institute for Mathematical Sciences,
 University of Edinburgh, Peter Guthrie Tait Road, Edinburgh EH9 3FD, UK}

\begin{abstract}

The asymptotic structure of gauge theories describing fracton interactions is analyzed. Two sets of asymptotic conditions are proposed. Both encompass all known solutions, lead to finite charges and resolve the problem of the divergent energy coming from the monopole contribution. While the first set leads to the expected fracton symmetry algebra, including a dipole charge, the second set provides a soft infinite-dimensional extension of it. These soft charges provide evidence of a rich infrared structure for fracton-like theories and provide one corner of a possible fracton infrared triangle.

\end{abstract}

\maketitle

\section{Introduction}
\label{sec:introduction}

Fractons~\cite{Chamon:2004lew,Haah:2011drr} are quasiparticles with
limited mobility and compose a novel, at this point theoretical, phase
of matter~\cite{Vijay:2015mka,Vijay:2016phm}. Their unusual properties
might be useful in the construction of quantum information
storage~\cite{Haah:2011drr,PhysRevLett.111.200501,RevModPhys.87.307,Brown:2019hxw}
and provide insights to a wide variety of physical fields, such as
quantum field theory~\cite{Seiberg:2020bhn,Seiberg:2020wsg} (and
follow-up works), general relativity~\cite{Xu_2006,Pretko:2017fbf},
elasticity~\cite{Pretko:2017kvd}, and even
holography~\cite{Yan:2018nco,Ganesan:2020wvm}. We refer to the reviews
\cite{Nandkishore:2018sel,Pretko:2020cko,Grosvenor:2021hkn} for
further applications, details, and references.

For some of these theories these remarkable properties can be traced
back to the existence of an electric charge $q$ and dipole charge
$\vec d$~\cite{Pretko:2016kxt,Pretko:2016lgv}
\begin{align*}
  q&=\int \rho \dd^{3} x & \vec d&= \int \vec x \rho \dd^{3}x  \, ,
\end{align*}
together with the conservation equation
$\dot \rho + \pd_{i}\pd_{j} J^{ij} = 0$. Using these conservation laws
one can infer that isolated charges are immobile, but dipoles can move
in restricted ways in accordance with dipole conservation. It follows
from the conservation equation that $\dot{\vec d}=0$. A single
particle with charge $e$ and trajectory $\vec\gamma(t)$, where
$\rho = e \delta(\vec x - \vec\gamma(t))$, is therefore restricted to
$\dot{\vec d}=\dot{\vec\gamma} = 0$, i.e., is immobile. For two
opposite charges where
$\rho=e \delta(\vec x-\vec\gamma_{1}(t)+\vec\gamma_{2}(t))$ dipole
conservation leads to $\dot{\vec{\gamma}}_{1}=\dot{\vec{\gamma}}_{2}$
which shows that they are allowed to move but only in prescribed ways.

The mediator of the interaction, the analog of the electromagnetic
field, is a gauge theory of rank two tensors given by an
electric(-like) field $E^{ij}$ and magnetic(-like) field
$B^{ij} = \epsilon^{imn}\pd_{m}A\indices{_{n}^{j}}$, see
\eqref{eq:action} for the action. The so called scalar charge gauge
theory~\cite{Pretko:2016kxt,Pretko:2016lgv} has a gauge symmetry
$\vd A_{ij} = \pd_{i}\pd_{j}\lambda$ and shares features with
electrodynamics (abelian gauge symmetry), general
relativity~\cite{Pretko:2017fbf} (rank two tensors) and partially
massless gravity~\cite{Bidussi:2021nmp} (higher derivative gauge
transformations). Its relation to microscopic lattice models is
reviewed in Appendix D of~\cite{Pretko:2016lgv}.

Motivated by recent advances in the understanding of the infrared
behavior of gauge theories (see~\cite{Strominger:2017zoo} for a
review), we will analyze the asymptotic structure, symmetries and
charges of this fracton gauge theory. This is a subtle endeavor since:
\begin{itemize}
\item The energy of an electric monopole, the most fundamental
  solution of the theory, is divergent as
  $E \sim \lim_{r \to \infty} r$~\cite{Pretko:2016lgv,Pretko:2016kxt}.
  This infinite energy is generic and a putative unphysical feature of
  this theory, that emerges for any charge distribution with nonzero
  total charge.
\item Due to the explicit $x^{i}$ factor the dipole charge $d^{i}$
  needs to handled with care. For large $x^{i}$ one needs to assure
  that the dipole charge remains finite and that its action on the
  fields preserves the allowed asymptotic conditions.\footnote{For
    further insightful remarks and an interesting complementary
    analysis of dipole charges, see~\cite{Gorantla:2022eem}.}
 \end{itemize}
We will show that the implementation of a consistent set of asymptotic
conditions, together with a careful treatment of the boundary terms,
resolves both of these issues.

For the well-definiteness of the theory, we propose asymptotic
conditions that encompass all known solutions and should be thought of
as describing arbitrary sources in a finite spatial region in space.
This can be interpreted as isolated systems as seen from far away and
means that we investigate the Coulumbic, rather than the radiative,
sector of the theory. Isolated systems are idealizations that allow to
discuss subsystems and assign to them physical attributes (like, e.g.,
energy, momentum, charge, dipole charge). For some theories this seems
trivial, like in Newtonian mechanics where isolated systems fall of
like $\frac{1}{r}$, however for gauge theories this is a subtle issue,
since some of the to-be-believed gauge redundancies might lead to
physical charges. This is indeed the case for electrodynamics, general
relativity and for the theories we are considering here. (For further
motivation see, e.g., Section 1 in~\cite{Geroch:1977jn} and
Section~2.10 in~\cite{Strominger:2017zoo}.)

We propose asymptotic conditions for which all charges are finite and
integrable and the asymptotic symmetries are indeed the expected
symmetries. The problem of the divergence in the energy is resolved
similarly to the infinitely long charged string in electrodynamics.
Thus, with these asymptotic conditions, the theory is well-defined and
possesses a finite action principle.

Besides the just described asymptotic conditions we propose a second
set which infinitely extends these symmetries. In analogy to
electrodynamics these novel charges can be thought of as soft charges.
This provides a hint towards a ``fracton infrared triangle'', which
apart from the asymptotic symmetries in one corner, should be
completed with a fracton memory effect and soft theorem. For a
discussion of infrared triangles we refer to the
review~\cite{Strominger:2017zoo}~(see also
\cite{Raclariu:2021zjz,Pasterski:2021rjz}) and references therein.

This work is structured as follows. Since the asymptotic analysis is
most easily done in spherical coordinates we discuss in
Section~\ref{sec:theory} the scalar charge gauge theory and its
symmetry transformations in curvilinear coordinates. Asymptotic
symmetries should encompass all interesting solutions, which are
reviewed in Section~\ref{sec:sol}. Our main results are summarized in
Section~\ref{sec:bound-cond-asympt}. As a first step we provide in
Section~\ref{sec:fracton-symmetries} a set of asymptotic conditions
which lead to a finite-dimensional symmetry algebra and resolve all
unphysical divergences of the energy and charges. In the second
part~\ref{sec:extended-fract-bound} we infinitely extend these
symmetries. We conclude with a discussion and outlook in
Section~\ref{sec:discussion}. As supplemental material we provide
details of the transformation of the fields in
Appendix~\ref{sec:suppl-mater}. Further details will be provided in a
future work~\cite{Perez:2022xxx}.

\section{Scalar charge gauge theory}
\label{sec:theory}

To understand the asymptotic structure of gauge theories with dipole
symmetry we focus on the arguably simplest example, the scalar charge
gauge theory~\cite{Pretko:2016kxt,Pretko:2016lgv}. We work on a fixed
flat non-lorentzian space-time in $3+1$ dimensions, parametrized by
$(t,x^{i})$, and with spatial slices that are given by noncompact
euclidean space $\R^{3}$. The action in Hamiltonian form for the symmetric
canonical variables $A_{ij}$ and $E^{ij}$ in curvilinear coordinates
is given by
\begin{align}
  \label{eq:action} 
  I[A_{ij},E^{ij},\phi]=\int dt d^{3}x\left(E^{ij}\dot{A}_{ij}-\mathcal{H}-\phi\nabla_{i}\nabla_{j}E^{ij}\right)\, ,
\end{align}
where the Hamiltonian $H$ is given by
\begin{align}
  \label{eq:H}
  H= \int d^{3} x\, \mathcal{H} = \int d^{3} x \frac{1}{2\sqrt{g}}
     \left( E_{ij}E^{ij} + B_{ij}B^{ij}\right) \, .
\end{align}
We have defined the ``magnetic'' field
$B^{ij} = \epsilon^{imn}\cd_{m}A\indices{_{n}^{j}}$ using the
Levi-Civita symbol $\epsilon^{ijk}$, so $B^{ij}$ as well as $E^{ij}$
are tensor densities of weight one. The spatial indices are lowered
and raised with the spatial metric $g_{ij}$ of which $g$ is the
determinant and $\cd_{i}$ the covariant derivative. We will often use
spherical coordinates
$g_{ij}dx^{i}dx^{j}=dr^{2}+r^{2}\gamma_{AB}dx^{A}dx^{B}$ where
$\gamma_{AB}$ is the metric of the round $2$-sphere, which we use to
raise and lower their respective indices. It has determinant $\gamma$
and covariant derivative $D_{A}$.

The action is invariant under the following transformations:
\begin{subequations}
  \label{eq:transf}
\begin{align}
  \delta A_{ij}&=
                 T\left(\frac{1}{\sqrt{g}}E_{ij}+\cd_{i}\cd_{j}\phi\right) +\cd_{i}\cd_{j}\lambda+\Ld_{\xi}A_{ij}
                  \\
  \delta E^{ij}&=
                 \frac{T}{2\sqrt{g}}\left(\epsilon\indices{^{mi}_{n}}\nabla_{m}B^{nj} + \epsilon\indices{^{mj}_{n}}\nabla_{m}B^{ni}\right)+\Ld_{\xi}E^{ij} \\
  \delta\phi&=\xi^{k}\partial_{k}\phi+\dot{\lambda}  \, , \label{eq:varphi}
\end{align}
\end{subequations}
where $\Ld$ denotes the Lie derivative. The time evolution is
parametrized by the constant $T$ and gauge transformations by
$\lambda(t,x^{i})$. They are generated by the Hamiltonian $H$ and the
constraint
\begin{align}
  \label{eq:Gauge}
  \mathscr{G}[\lambda] = \int d^{3}x \lambda \cd_{i}\cd_{j}E^{ij}
 \, ,
\end{align}
respectively. In this section we will not consider boundary terms, but
they will be included in Section~\ref{sec:bound-cond-asympt}. The
spatial translations $\vec \alpha$ and spatial rotations $\vec \omega$
are parametrized by a vector $\xi^{i}$ that obeys
$\nabla_{i}\xi_{j}+\nabla_{j}\xi_{i} = 0$. In spherical coordinates
they are given by
\begin{align*}
  \xi^{r} & =\vec{\alpha}\cdot\hat{r} &
 \xi^{A} & =\frac{\epsilon^{AB}}{\sqrt{\gamma}}\partial_{B}\left(\vec{\omega}\cdot \hat{r}\right)
            +\frac{1}{r}\partial^{A}\left(\vec{\alpha}\cdot\hat{r}\right)
\end{align*}
($\hat{r}$ is the unit radial vector) and can be canonically
generated via
\begin{align*}
  G\left[\xi^{i}\right]
  =\int d^{3}x\,
  \left(\xi^{k}E^{ij}\nabla_{k}A_{ij}+2E^{ij}A_{kj}\nabla_{i}\xi^{k}\right) \, .
\end{align*}
This generator is not manifestly gauge invariant due to the explicit
dependence on $A_{ij}$, cf.,~\eqref{eq:transf}. However, its on-shell
value (more precisely its value on the constraint surface) is not
affected by gauge transformations since the variations are
proportional to the Gauss constraint $ \cd_{i}\cd_{j}E^{ij} = 0$.
Therefore, momentum and angular momentum are physical observables.

In electrodynamics it is possible to define ``improved generators''
that are manifestly gauge invariant by adding a specific
field-dependent gauge transformation (improved energy-momentum
tensor). In the case of the momentum, this leads to the Poynting
vector. For this theory, the analog of the Poynting vector would be
$\mathcal{S}_{k} = E^{ij}(\pd_{k}A_{ij} - \pd_{i}A_{kj})$ (in
Cartesian coordinates). This quantity, that appears in the continuity
equation for the energy $\dot{\mathcal{H}} = \pd_{k} \mathcal{S}^{k}$,
is however not conserved. It is not possible to improve the momentum
with a local gauge transformation to obtain
$\mathcal{S}_{k}$~\cite{Jain:2021ibh}, let us however remark that it
can be accomplished using a non-local gauge transformation of the
form~\eqref{eq:Gauge} with $ \pd_{i}\lambda = -\xi^{j} A_{ij}$.

One of the key properties of this theory is that it possesses no boost
symmetry that mixes space and time, more precisely, this theory lives
on a flat Aristotelian
geometry~\cite{Figueroa-OFarrill:2018ilb,Bidussi:2021nmp,Jain:2021ibh}.

\section{Solution space}
\label{sec:sol}

All the physically relevant solutions of the theory must be part of a
consistent set of asymptotic conditions, and guide the asymptotic
fall-offs that one should impose, e.g., boundary conditions for
electrodynamics should accommodate the Coulomb solution. We summarize
the for our use relevant solutions of the scalar charge
theory~\cite{Pretko:2016kxt}.

A particular feature of this theory is that solutions sourced by
isolated point charged particles have a slower radial fall-off than
for example in electrodynamics. This is the root of the divergence in
the energy. Furthermore, the dipole solutions play a fundamental role
for the asymptotic analysis due to the presence of conserved dipole
charges in this theory.

We start with the simplest solution, an isolated static point particle
with charge $e$ sitting at the origin. In spherical coordinates the
nonzero fields for this electric monopole are given by
\begin{align*}
  E_{\text{mono}}^{AB}&=\frac{\sqrt{\gamma}\gamma^{AB}}{8\pi}\frac{e}{r} & \phi_{\text{mono}}&=-\frac{e}{8\pi} r \, .
\end{align*}
A particular property of the electric monopole is that the energy is
linearly divergent. Indeed, as it was pointed out
in~\cite{Pretko:2016kxt}, for large values of $r$ the energy is given
by $E=\frac{e^{2}}{16\pi}r + \text{finite terms}$. We will show that,
with a careful treatment of boundary terms, this potentially
unphysical situation can be resolved.

Another solution of interest is the (pure ideal) electric dipole
$\vec p$ with non-vanishing fields given by
\begin{align*}
  E_{\text{dip}}^{rA}&=\frac{\sqrt{\gamma}}{8\pi} \frac{\partial^{A}\left(\vec{p}\cdot\hat{r}\right)}{r}
                       &
  &E_{\text{dip}}^{AB}=\frac{\sqrt{\gamma}\gamma^{AB}}{8\pi} \frac{\vec{p}\cdot\hat{r}}{r^{2}}  \\
   \phi_{\text{dip}}&=\frac{1}{8\pi} \vec{p}\cdot\hat{r} \, .
\end{align*}
As expected, the fall-offs are subleading with respect to the electric
monopole, but still strong enough to lead to non-vanishing charges.

The solution corresponding to a ``magnetic particle'' is given by
\begin{align*}
 B^{rr}&=\frac{\sqrt{\gamma}}{8\pi} \vec{m}\cdot\hat{r} &
 B^{Ar}&=\frac{\sqrt{\gamma}}{16\pi} \frac{\partial^{A}  \left( \vec{m}\cdot\hat{r} \right)}{r}  \\
 B^{AB}&=-\frac{\sqrt{\gamma}\gamma^{AB}}{16\pi} \frac{\vec{m}\cdot\hat{r}}{r^{2}} &
 B^{rA}&=\frac{5\sqrt{\gamma}}{16\pi} \frac{\partial^{A} \left( \vec{m}\cdot\hat{r} \right)}{r} \, .
\end{align*}
In the gauge where the linear term in $r$ of $\gamma^{AB}A_{AB}$
vanishes, the nonzero components of the potential take the form
\begin{align*}
  A_{rA}&=\frac{\sqrt{\gamma}}{16 \pi}\epsilon_{AB}\partial^{B}\left(\vec{m}\cdot\hat{r}\right) \\
  A_{AB}&=\frac{\sqrt{\gamma}}{16 \pi r}\left[\epsilon_{AC}D^{C}D_{B}+\epsilon_{BC}D^{C}D_{A}\right]\left(\vec{m}\cdot\hat{r}\right)\,.
\end{align*}

\section{Asymptotic conditions and symmetries}
\label{sec:bound-cond-asympt}

Asymptotic conditions describe the behavior of the fields near
infinity and are of fundamental importance to determine physical
symmetries of gauge theories. There is in general no unique set of
asymptotic conditions\footnote{According to
  Geroch~\cite{Geroch:1977jn}:\emph{There are no ``correct'' or
    ``incorrect'' definitions, only more or less useful ones. It is
    perfectly possible that there turn out to be a number of competing
    definitions, applicable to differing physical systems, or a single
    definition as in Newtonian gravitation, or none at all.}} however
the following physical requirements must be fulfilled:
\begin{itemize}
\item The conditions should encompass all relevant physical solutions,
  in particular linear combinations of the ones described
  in~\cite{Pretko:2016kxt} and reviewed in~Section~\ref{sec:sol}.
\item The charges (energy, momentum, angular momentum, electric and
  dipole) and the symplectic structure must be finite, which
  guarantees that the action is also finite.
\end{itemize}
As it will be seen below, this puts severe restrictions on our theory.
We propose two sets of asymptotic conditions that remarkably satisfy
all these consistency requirements. While the first set reproduces the
expected finite-dimensional fracton symmetries, the second set extends
the symmetry algebra to an infinite-dimensional one containing novel
``soft charges.''

\subsection{Fracton symmetries}
\label{sec:fracton-symmetries}

\subsubsection{Asymptotic conditions}
\label{sec:asympt-cond}

\begingroup
\allowdisplaybreaks

The asymptotic conditions leading to a finite-dimensional symmetry
algebra are given by
\begin{subequations}
  \label{eq:conv-exp}
\begin{align}
  E^{rr} &= E^{rr}_{(0)}+O\left(r^{-1}\right)  \label{eq:Err} \\
  E^{rA} &= \frac{E_{\left(-1\right)}^{rA}}{r}+O\left(r^{-2}\right) \\
  E^{AB} &= \frac{\sqrt{\gamma}\gamma^{AB}}{8\pi}\frac{q}{r}+\frac{E_{\left(-2\right)}^{AB}}{r^{2}}+O\left(r^{-3}\right) \label{eq:EAB}\\
  A_{rr} &= \frac{A_{rr}^{(-1)}}{r}+O\left(r^{-2}\right) \\
  A_{rA} &= A_{rA}^{\left(0\right)}+O\left(r^{-1} \right) \\
  A_{AB} &= A_{AB}^{\left(1\right)}r+A_{AB}^{\left(0\right)}+O\left(r^{-1}\right)\, , \quad\,  \gamma^{AB}A^{(1)}_{AB}=0 \label{eq:AAB-bc}\\
 \phi &=  \left( \vec{\Phi}^{\left(1\right)}\cdot\hat{r} -\frac{q}{8\pi} \right)r+\Phi^{\left(0\right)}+O\left(r^{-1}\right) \label{eq:phi-bc}
\end{align}
and are preserved under time evolution, rotations, translations, and
gauge transformations with the following parameter
\begin{align}
    \lambda &=(\vec{\lambda}^{\left(1\right)}\cdot\hat{r}) \, r+\lambda^{\left(0\right)}+O\left(r^{-1}\right) \, . \label{eq:lambda-1} 
\end{align}
\end{subequations}
\endgroup Here, $q$ (the sum of all monopole charges),
$\Phi^{\left(0\right)}$, $\lambda^{\left(0\right)}$,
$\vec{\Phi}^{\left(1\right)}$ and $\vec{\lambda}^{\left(1\right)}$ are
constants with respect to the angles. The remaining terms are
functions on the sphere and have to satisfy the following parity
conditions
\begin{align*}
&E_{\left(-1\right)}^{r\theta},\,E_{\left(-2\right)}^{\theta\phi},\,A_{r\phi}^{\left(0\right)},\,A_{\theta\theta}^{\left(1\right)},\,A_{\phi\phi}^{\left(1\right)} &&\text{parity even} \\
&E_{\left(-1\right)}^{r\phi},\,E_{\left(-2\right)}^{\theta\theta},\,E_{\left(-2\right)}^{\phi\phi},\,A_{r\theta}^{\left(0\right)},\,A_{\theta\phi}^{\left(1\right)},\,A^{\left(0\right)} &&\text{parity odd}
\end{align*}
where we denote traces with respect to the sphere, like $A^{(0)}$, as
$X=\gamma^{AB}X_{AB}$. The tracefree part of $A_{AB}^{(0)}$ is
unconstrained. The parity conditions for the fields $E^{rr}_{(0)}$ and
$A_{rr}^{(-1)}$ cannot be inferred from the known solutions. However,
in order to guarantee a finite symplectic term, they must have
opposite parity, or at least one of them must vanish. All of these
conditions are fully consistent with the preservation of the
asymptotic symmetries.

The form of the asymptotic conditions in~\eqref{eq:conv-exp} guarantee
that the charges and the action principle do not possess divergences
in the large $r$ limit. For example, the leading term of $E^{AB}$ in
\eqref{eq:EAB} must only have a trace part in order to cancel the
linear divergence appearing in the bulk Hamiltonian, with the one
coming from the boundary term of the Gauss constraint. In particular,
the shift that was done in the leading order of $\phi$ in
\eqref{eq:phi-bc} is of fundamental importance for this purpose, as it
will be explained in detail below. This shift is also necessary to
accommodate the monopole solution within the asymptotic expansion.
Note that a constant $q$ is compatible with the leading term of the
Gauss constraint, $D^{A}D_{A}q=0$, as it must.

The condition $A^{(1)}=0$ in \eqref{eq:AAB-bc} removes a linear
and a logarithmic divergence in the symplectic term. The preservation
of this condition restricts the leading orders of $\phi$ and $\lambda$
to take the form exhibited in \eqref{eq:phi-bc} and
\eqref{eq:lambda-1}. In the next order, $\lambda^{(0)}$ and
$\Phi^{(0)}$ could in principle have higher modes in the spherical
harmonic expansion, however the modes with $\ell \geq 1$ do not appear
neither in the charges nor in the boundary term of the action
principle. Therefore they are ``pure gauge'' and can be consistently
discarded.

The parity conditions are necessary to remove additional logarithmic
divergences appearing in the symplectic term, as well as in the
boundary terms associated with the Gauss constraint. They have a long
history in the Hamiltonian formulation of general relativity and
electrodynamics~\cite{Regge:1974zd,Henneaux:2018cst,Henneaux:2018gfi}
and it should thus not come as a surprise that they are also needed
for fracton theories.

\subsubsection{Conserved charges}
\label{sec:conserved-charges}

The charges associated with gauge transformations are obtained from
the boundary term of the Gauss constraint~\cite{Regge:1974zd} (see
also~\cite{Benguria:1976in}) and are given by
\begin{align*}
Q=  \int d^{3} x \pd_{i}\left(\partial_{j}\lambda \, E^{ij}-\lambda\nabla_{j} E^{ij}\right)= \lambda^{\left(0\right)}q+\vec{\lambda}^{\left(1\right)}\cdot\vec{d},
\end{align*}
where $q$ corresponds to the total electric charge and 
\begin{align*}
  \label{eq:di-charge}
  \vec{d}=\oint d^{2}x\,\hat{r}\left(E^{rr}_{(0)} + E_{\left(-2\right)} - 2D_{A}E_{\left(-1\right)}^{rA} \right)\,,
\end{align*}
corresponds to the total dipole charge. As expected for gauge
symmetries the charges are surface terms integrated at infinity.

\subsubsection{Finiteness of the energy}
\label{sec:well-defined-energy}

As was explained previously, the energy~\eqref{eq:H} diverges linearly
in $r$ for the monopole solution~\cite{Pretko:2016lgv,Pretko:2016kxt}.
This is also true for the asymptotic conditions. Indeed, if we
evaluate the Hamiltonian $H$ in~\eqref{eq:H} using our boundary
conditions~\eqref{eq:conv-exp}, the following divergence is obtained
\begin{equation}
  \label{divH}
  H=\lim_{r \to \infty} r\oint d^{2}x\,\sqrt{\gamma}\left(\frac{q}{8\pi}\right)^{2} + \text{ finite terms} \, .
\end{equation}
We will show that a careful treatment of the boundary terms in the
total Hamiltonian~\cite{Dirac:1950pj,dirac2001lectures2}, completely
removes the divergence for any physical configuration that fulfills
our asymptotic conditions.

Let us consider the total Hamiltonian that includes the constraints,
in the sense introduced by
Dirac~\cite{Dirac:1950pj,dirac2001lectures2},
\begin{align}
  \label{eq:Htot}
  H_{T}=H+\int d^{3}x\,\phi \nabla_{i}\nabla_{j}E^{ij}+B_{\infty} \, .
\end{align}
Here $B_{\infty}$ is the boundary term needed to guarantee that the
generator has well-defined functional derivatives~\cite{Regge:1974zd}
and has a variation of the form
\begin{align*}
  \delta B_{\infty}=\int d^{3} x \pd_{i}\left( \partial_{j}\phi \, \delta E^{ij}-\phi\nabla_{j} \delta E^{ij}\right).
\end{align*}
Using~\eqref{eq:conv-exp} and by virtue of the shift in the leading
order of $\phi$ in~\eqref{eq:phi-bc}, this expression can be
integrated in field space and acquires a linear divergence given by
\begin{align*}
  B_{\infty}=- \lim_{r \to \infty} r\oint d^{2}x\,\sqrt{\gamma}\left(\frac{q}{8\pi}\right)^{2} \, .
\end{align*}
In the total Hamiltonian~\eqref{eq:Htot}, $B_{\infty}$ precisely
cancels the divergence coming from $H$, c.f.,~\eqref{divH}. Therefore
the total Hamiltonian is finite and provides a well-defined notion of
energy for any distribution of charges.\footnote{A similar method was
  used in~\cite{Perez:2015kea, Perez:2015jxn} to regularize the energy
  of the charged black hole in three-dimensional gravity.}

Thus, the final expression for the finite energy takes the following
form
\begin{align}
  \label{eq:H-finite}
  E_{\text{finite}}= H- \lim_{r \to \infty}  r\oint d^{2}x\,\sqrt{\gamma}\left(\frac{q}{8\pi}\right)^{2}.
\end{align}

\subsubsection{Fracton symmetry algebra}
\label{sec:symmetry-algebra}

The symmetry algebra can be obtained directly from the Dirac brackets
and is spanned by the generators of rotations $J_{I}$, translations
$P_{I}$ (that do not need to be improved by boundary terms), the
energy $E_{\mathrm{finite}}$ which is a trivial central extension, and
by the generators $q$ and $d_{I}$ that correspond to the electric
charge and dipole moment and are associated with the large gauge
symmetries. The non-vanishing commutators are given by
\begin{subequations}
  \label{eq:asymp-conv}
  \begin{align}
    \left\{ J_{I},J_{J}\right\} &=\epsilon_{IJK}J_{K} & \left\{ J_{I},P_{J}\right\} &=\epsilon_{IJK}P_{K} \\
    \left\{ J_{I},d_{J}\right\} &=\epsilon_{IJK}d_{K} & \left\{ P_{I},d_{J}\right\} &=\delta_{IJ} q \,. 
\end{align}
\end{subequations}
Here $I,J,K=1,2,3$ denote the Cartesian components of the generators.
For a theory with conserved dipole moment this is precisely the
algebra one expects. Let us emphasize that without the regularized
energy this symmetry algebra would not be well defined.

\subsection{Extended fracton symmetries}
\label{sec:extended-fract-bound}

In recent years, deep relations between the asymptotic structure of
gauge theories and their infrared behavior have been highlighted (see,
e.g.,~\cite{Strominger:2017zoo} for a review). One of the cornerstones
of this relation is the existence of infinite-dimensional symmetries
that contain additional ``soft charges.'' Therefore, one might wonder
if there exists an infinite-dimensional extension of the fracton
symmetries~\eqref{eq:asymp-conv}, similar to the extension of the
Poincaré algebra, to the Bondi-Metzner-Sachs
algebra~\cite{Bondi:1962px,Sachs:1962zza}. Inspired
by~\cite{Henneaux:2018cst,Henneaux:2018gfi,Henneaux:2018hdj} we
construct an alternative set of asymptotic conditions that allows such
extension.

\subsubsection{Asymptotic conditions}
\label{sec:asympt-cond-1}

With respect to the asymptotic conditions \eqref{eq:conv-exp} we adapt
the following lines
\begin{subequations}
  \label{eq:asymp-cond-ext}
\begin{align}
  E^{AB} & =\frac{\sqrt{\gamma}\gamma^{AB}}{8\pi}\frac{q}{r}+\frac{\sqrt{\gamma}\gamma^{AB}}{8\pi} \frac{\vec{p}\cdot\hat{r}}{r^{2}}+O\left(r^{-3}\right) \label{eq:EABext}\\
  A_{AB} &= A_{AB}^{\left(1\right)}r+A_{AB}^{\left(0\right)}+O\left(r^{-1}\right) \\
  \phi&= \left(\Phi^{\left(1\right)} - \frac{q}{8\pi} \right)r+\Phi^{\left(0\right)}+O\left(r^{-1}\right) \\
  \lambda&= \lambda^{\left(1\right)}r+\lambda^{\left(0\right)}+O\left(r^{-1}\right) \label{eq:lambda-2}  
\end{align}
\end{subequations}
where $q$, $\vec{p}$ are constant with respect to the angles.
Additionally, we have to impose the following spherical dependence
($\ell$ denotes the degree of the spherical harmonic function
$Y_{\ell,m}(\theta,\phi)$)
\begin{align*}
  &A^{(0)},\Phi^{\left(1\right)},\lambda^{\left(1\right)} &&\ell \geq 1 \\
  &A^{(1)} &&\ell \geq 2 \\
  &E_{\left(-1\right)}^{r\theta},\,A_{r\phi}^{\left(0\right)} &&\text{parity even} \\
  &E_{\left(-1\right)}^{r\phi}\,A_{r\theta}^{\left(0\right)} &&\text{parity odd} \, .
\end{align*}
We set $\Phi^{\left(0\right)}$ and $\lambda^{\left(0\right)}$ to be
constant, they could in principle have higher modes which are however
pure gauge. As in the previous case, $E^{rr}_{(0)}$ and
$A_{rr}^{(-1)}$ must have opposite parity, or at least one of them
must vanish. With the above conditions, the symplectic term and the
charges are finite. Furthermore, the divergence in the energy coming
from the bulk Hamiltonian is removed exactly like
in~\eqref{eq:H-finite}.

\subsubsection{Conserved charges}
\label{sec:conserved-charges-1}

If we expand the parameter $\lambda^{(1)}$ in spherical harmonics
\begin{align}
\lambda^{\left(1\right)}=\vec{\lambda}^{\left(1\right)}\cdot\hat{r}+\sum_{\ell\geq2}\sum_{m=-\ell}^{\ell}\lambda_{\ell,m}Y_{\ell,m}
\end{align}
then the charge associated with gauge transformations takes the form
\begin{align*}
  Q &= \lambda^{\left(0\right)}q + \vec{\lambda}^{\left(1\right)}\cdot\vec{d} + \sum_{\ell\geq2}\sum_{m=-\ell}^{\ell}\lambda_{\ell,m}Q_{\ell,m}
\end{align*}
where
\begin{align*}
\vec{d} & =\frac{1}{3}\vec{p}+\oint d^{2}x\,\hat{r}\left( E^{rr}_{(0)} - 2D_{A}E_{\left(-1\right)}^{rA}\right) \\
Q_{\ell,m} & =\oint d^{2}x\,Y_{\ell,m}\left(E^{rr}_{(0)} - 2D_{A}E_{\left(-1\right)}^{rA}\right) \, .
\end{align*}
As in the previous case, $q$ and $\vec{d}$ correspond to the electric
and dipole charges, respectively. In addition, there is an infinite
tower of new charges characterized by multipoles with $\ell \geq 2$.
We call them ``soft charges'' by analogy with the infinite-dimensional
extension of the asymptotic symmetry algebra in electrodynamics and
general relativity.

\subsubsection{Extended fracton symmetry algebra}
\label{sec:extend-fract-symm}

The symmetry algebra is given by the non-vanishing Poisson
brackets~\eqref{eq:asymp-conv}, together with the extension
\begin{align}
  \label{eq:asymp-ext}
  \left\{ J_{I},Q_{\ell,m}\right\} &=\sum_{m'=-\ell}^{\ell} (D_{mm'}^{\ell})_{I}Q_{\ell,m'} \quad (\ell \geq 2, |m| \leq \ell)
\end{align}
where $D_{mm'}^{\ell}$ is the Wigner $D$-matrix that rotates the
spherical harmonics. Note that the soft charges commute with the
translation generator, in contrast to the dipole charge. This property
can be seen as the imprint of their soft nature.\footnote{For
  fracton-like theories the conserved higher multipole charges
  generically do not commute with the translations (see,
  e.g.,~\cite{Gromov:2018nbv}) and are therefore not soft.}

\section{Discussion and outlook}
\label{sec:discussion}

This work provides the first asymptotic analysis of a theory with
conserved dipole charge and suggests the existence of a rich infrared
structure. The first set of asymptotic conditions, provided in
Section~\ref{sec:asympt-cond} leads to the expected finite-dimensional
fracton algebra~\eqref{eq:asymp-conv}. The second set of asymptotic
conditions, see Section~\ref{sec:asympt-cond-1}, provides an
infinite-dimensional ``soft'' extension of the dipole charges,
cf.,~\eqref{eq:asymp-ext}. In both cases a careful analysis of the
boundary terms was of fundamental importance in order to obtain a
finite energy and a well-defined action principle.

In a subtle way the extended asymptotic
conditions~\eqref{eq:asymp-cond-ext} also encompass the more
restricted ones~\eqref{eq:conv-exp}. If in~\eqref{eq:asymp-cond-ext}
the condition $A^{(1)}=0$ is imposed, the algebra truncates to the
finite-dimensional one in \eqref{eq:asymp-conv}.
Comparing~\eqref{eq:EAB} with~\eqref{eq:EABext} it seems that that
there is more freedom in the restricted asymptotic conditions. This
freedom is however irrelevant for the asymptotic symmetries since only
the trace part of $E^{AB}_{(-2)}$ contributes to the charges.

We have seen that the electric charge, in contradistinction to the
dipole charge, stays unextended. This is related to the requirement of
finite energy from which the specific form of the first term on the
right hand side of~\eqref{eq:EAB} follows. Together with the
constraint of the theory this leads to a charge that is independent of
the angles. Dropping these restrictions opens the possibility to also
extend the charge sector (at the prize of an infinite energy).

The present work opens various avenues for further research. One is
the generalization of this analysis to other interesting models like
the traceless scalar charge theory, their vector generalizations and
beyond, see,
e.g.,~\cite{Pretko:2016kxt,Pretko:2016lgv,Bulmash:2018knk,Schmitz:2018kbo,Slagle:2018swq,Gromov:2018nbv,Prem:2019etl,Radzihovsky:2019jdo,Li:2019tje,Wang:2019aiq,Shenoy:2019wng,Radicevic:2019vyb,Wang:2019cbj,Seiberg:2019vrp,Argurio:2021opr,Casalbuoni:2021fel,Pena-Benitez:2021ipo,Angus:2021jvm}.
Another possibility is to allow for curved
space(-time)~\cite{Slagle:2018kqf,Bidussi:2021nmp,Jain:2021ibh}. It
would also be interesting to study the consequences of physical
boundaries at finite distances and their effect on the symmetries of
the system. In each case new phenomena and features should emerge.

We have focused on the Coulumbic sector, but this theory also allows
for radiative modes that could make it possible to connect the soft
charges, possibly via vacuum transitions, to a putative fracton memory
effect. This is one way to argue for the measurable consequences of
the novel soft charges. Besides the discussed asymptotic symmetries,
the memory effect and soft theorems could provide the three corners of
a novel fracton infrared triangle that remains to be explored and
could lead to interesting applications in condensed matter and high
energy physics.

\begin{acknowledgments}
  We want to thank Kristan Jensen for useful discussions. SP is
  grateful to Leo Bidussi, Jelle Hartong, Emil Have and Jørgen Musaeus
  for collaboration on aspects of fractonic theories. The research of
  AP is partially supported by Fondecyt grants No 1211226, 1220910. SP
  is supported by the Leverhulme Trust Research Project Grant
  (RPG-2019-218) ``What is Non-Relativistic Quantum Gravity and is it
  Holographic?''. We would like to thank the organizers of the Carroll
  Workshop in Vienna where part of this work was completed.
\end{acknowledgments}

\appendix

\section{Transformations of the fields}
\label{sec:suppl-mater}

\begingroup
\allowdisplaybreaks

The transformations of the expanded fields are,
using
$Y^{A} =\frac{\epsilon^{AB}}{\sqrt{\gamma}}\partial_{B}\left(\vec{\omega}\cdot
  \hat{r}\right)$ and $\Delta=D_{A}D^{A}$,
given by
\begin{subequations}
\begin{align*}
  \delta q&=0 \\
\delta E_{\left(0\right)}^{rr} &=\mathcal{L}_{Y}E_{\left(0\right)}^{rr} \\
\delta E_{\left(-1\right)}^{rA} &=\mathcal{L}_{Y}E_{\left(-1\right)}^{rA}-\frac{q}{8\pi}\sqrt{\gamma}\partial^{A}\left(\vec{\alpha}\cdot\hat{r}\right) \\
\delta\tilde{E}_{\left(-2\right)}^{AB} &=\mathcal{L}_{Y}\tilde{E}_{\left(-2\right)}^{AB} \\
\delta E_{\left(-2\right)} &=\mathcal{L}_{Y}E_{\left(-2\right)}-\frac{q}{4\pi}\sqrt{\gamma} \left(\vec{\alpha}\cdot\hat{r}\right)
\end{align*}
\begin{align*}
  \delta A^{\left(1\right)} &=\mathcal{L}_{Y}A^{\left(1\right)}+T\left(\Delta+2\right)\Phi^{\left(1\right)}+\left(\Delta+2\right)\lambda^{\left(1\right)} \\
  \delta A^{\left(0\right)}  &=\mathcal{L}_{Y}A^{\left(0\right)} + 2 \partial^{A} \left(\vec{\alpha}\cdot\hat{r}\right)A_{rA}^{\left(0\right)} \nonumber \\
                            & \quad + \partial^{A}\left(\vec{\alpha}\cdot\hat{r}\right)\partial_{A}A^{\left(1\right)}-\left(\vec{\alpha}\cdot\hat{r}\right)A^{\left(1\right)} \nonumber \\
                            & \quad  + T\left(\Delta\Phi^{\left(0\right)}+\frac{1}{\sqrt{\gamma}}E_{\left(-2\right)}\right)+\Delta\lambda^{\left(0\right)} \\
  \delta A_{AB}^{\left(1\right)} &= \mathcal{L}_{Y}A_{AB}^{\left(1\right)} +T\left(D_{A}D_{B} -\frac{1}{2}\gamma_{AB}\Delta\right)\Phi^{\left(1\right)} \nonumber \\
                                           &\quad +\left(D_{A}D_{B} -\frac{1}{2}\gamma_{AB}\Delta\right)\lambda^{\left(1\right)} \\
  \delta A_{rr}^{\left(-1\right)} &=\mathcal{L}_{Y}A_{rr}^{\left(-1\right)} \\
\delta A_{rA}^{\left(0\right)} &=\mathcal{L}_{Y}A_{rA}^{\left(0\right)}  \, .
\end{align*}
\end{subequations}
\endgroup

\providecommand{\href}[2]{#2}\begingroup\raggedright\endgroup

\end{document}